\documentclass[conference]{IEEEtran}
\IEEEoverridecommandlockouts

\usepackage{cite}
\usepackage{amsmath,amssymb,amsfonts}
\usepackage{algorithmic}
\usepackage{graphicx}
\usepackage{textcomp}
\usepackage{xcolor}
\usepackage{comment}
\usepackage{bm}
\usepackage[utf8]{inputenc}
\usepackage[english]{babel}
\usepackage[T1]{fontenc}
\usepackage{amsthm,epsfig,color,empheq,graphics,algorithm,algorithmic,braket,dsfont,balance,accents,multirow}

\newtheorem{lemma}{Lemma}

\newcommand \bzero{\mathbf{0}}

\newcommand \bb{\mathbf{b}}

\newcommand \be{\mathbf{e}}

\newcommand \bs{\mathbf{s}}

\newcommand \bv{\mathbf{v}}

\newcommand \bx{\mathbf{x}}
\newcommand \by{\mathbf{y}}

\newcommand \bG{\mathbf{G}}
\newcommand \bH{\mathbf{H}}
\newcommand \bI{\mathbf{I}}

\newcommand \bR{\mathbf{R}}

\newcommand \bU{\mathbf{U}}
\newcommand \bV{\mathbf{V}}
\newcommand \bW{\mathbf{W}}
\newcommand \bX{\mathbf{X}}
\newcommand \bY{\mathbf{Y}}

\newcommand \btheta{\boldsymbol{\theta}}

\newcommand \bpsi{\boldsymbol{\psi}}

\newcommand \bLambda{\mathbf{\Lambda}}

\newcommand \mcH{\mathcal{H}}

\newcommand \mcO{\mathcal{O}}

\newcommand \tbG{\tilde{\mathbf{G}}}

\newcommand \tbpsi{\tilde{\boldsymbol{\psi}}}

\newcommand \hbb{\hat{\mathbf{b}}}

\def\BibTeX{{\rm B\kern-.05em{\sc i\kern-.025em b}\kern-.08em
    T\kern-.1667em\lower.7ex\hbox{E}\kern-.125emX}}

\begin{document}

\title{Learning AC Power Flow Solutions using a Data-Dependent Variational Quantum Circuit\\
\thanks{This work has been funded by the US National Science Foundation under grant 2412947, and the Office of Naval Research under grant N000142412614.}
}

\author{
\IEEEauthorblockN{Thinh Viet Le, Md Obaidur Rahman, and Vassilis Kekatos}
\IEEEauthorblockA{Elmore Family School of Electrical and Computer Engineering, Purdue University\\
West Lafayette, IN 47906, USA\\
Emails: \{le272,rahma160,kekatos\}@purdue.edu}
}

\maketitle

\begin{abstract}
Interconnection studies require solving numerous instances of the AC load or power flow (AC PF) problem to simulate diverse scenarios as power systems navigate the ongoing energy transition. To expedite such studies, this work leverages recent advances in quantum computing to find or predict AC PF solutions using a variational quantum circuit (VQC). VQCs are trainable models that run on modern-day noisy intermediate-scale quantum (NISQ) hardware to accomplish elaborate optimization and machine learning (ML) tasks. Our first contribution is to pose a single instance of the AC PF as a nonlinear least-squares fit over the VQC trainable parameters (weights) and solve it using a hybrid classical/quantum computing approach. The second contribution is to feed PF specifications as features into a data-embedded VQC and train the resultant quantum ML (QML) model to predict general PF solutions. The third contribution is to develop a novel protocol to efficiently measure AC-PF quantum observables by exploiting the graph structure of a power network. Preliminary numerical tests indicate that the proposed VQC models attain enhanced prediction performance over a deep neural network despite using much fewer weights. The proposed quantum AC-PF framework sets the foundations for addressing more elaborate grid tasks via quantum computing. 
\end{abstract}


\section{Introduction}
Solving the AC power flow (AC PF) is an imperative task in power system simulation, operation, and planning. Given specified load conditions and generator settings, the AC PF seeks the complex voltage phasors at all buses~\cite{tinney1967power}. Once voltage phasors are found, other quantities of interest, such as line power and currents, can be readily computed. Under the forthcoming energy transition, the increased uncertainty in generation, load demand, and prevailing technologies calls for conducting interconnection studies involving PF instances of ever-increasing volume, variability, and spatiotemporal resolution. This work explores recent advances in quantum machine learning (QML) to expedite bulk PF studies.

Traditional algorithms for solving the AC PF rely on linearization techniques or iterative nonlinear equation solvers~\cite{ExpConCanBook}. Linearized variants of the PF fail to capture line flows and thermal losses. Iterative methods coping with the AC PF equations include the Gauss-Seidel, the impedance matrix, and Newton-Raphson methods~\cite{7926417}. These tools have been studied extensively, and their mature implementations are widely used in industry practice. Reference~\cite{Madani2019} reformulates the AC PF task as a feasibility problem and relaxes it into a semidefinite program (SDP) to find AC-PF solutions even when Newton-Raphson iterates fail to converge. 

Previous quantum computing attempts to cope with the AC PF task have focused on the classical Newton-Raphson iterations. The idea was to merely substitute the linear system solver with the Harrow–Hassidim–Lloyd (HHL) algorithm~\cite{ saevarsson2022quantum,feng2021quantum,liu2024quantum}. The HHL algorithm runs on fault-tolerant rather than noisy intermediate-scale quantum (NISQ) quantum computers and can solve systems of $N$ linear equations in $\mathcal{O}(\log N)$. Nonetheless, its complexity scales unfavorably for linear systems involving the Jacobian of the AC-PF equations~\cite{demystifying}. 

All previous solvers deal with a single PF instance at a time. The quasi-static time series (QSTS) method resorts to elaborate initializations and sampling schemes for solving the AC PF in distribution grids across successive instances~\cite{reno2017motivation}. Similar ideas may not scale equally well under the wider variability of planning studies in transmission systems. To address this challenge, a recent line of research advocates training machine learning (ML) models, including graph and physics-informed neural networks, to predict AC-PF solutions once presented with the problem specifications~\cite{hu2021, gou2022,gao2024,jordan2024}. The motivation is to shift the computational burden to offline and perform the time-efficient prediction step during real-time operation. To sufficiently represent various AC-PF instances, the number of weights in classical ML models is massive, hindering both training and inference~\cite {jordan2024}. Moreover, previous ML models are typically trained in a supervised manner, requiring large datasets of labeled solved PF instances. 


Quantum ML (QML) models have been recently proposed as an alternative to classical ones~\cite{havlivcek2019supervised,abbas2021power}. However, available quantum hardware is confined in terms of width (number of qubits), depth (number of gate layers), and sensitivity to noise. Under these limitations, current QML models are designed to run on \emph{variational quantum circuits} (VQC)~\cite[Ch.~5]{schuld2021}. In a VQC, both inputs and trainable weights are entered as tunable parameters of controllable quantum gates, mathematically represented as parameterized unitary matrices. The VQC output is measured as the expectation over a \emph{quantum observable}, described by a quadratic function of the VQC state defined over a Hermitian matrix. Interestingly, recent results show that for the same number of trainable weights, a QNN can approximate smooth functions at higher accuracy than any classical NN~\cite{abbas2021power,yu2024}. 

Measuring quantum observables is indispensable during both training and inference of a VQC. The VQC output can be measured using the technique of linear combination of unitaries (LCU)~\cite[p.~138]{schuld2021}. Unfortunately, for problems featuring Hermitian matrices with dense diagonals, the number of unitaries scales exponentially with the number of qubits. Reference~\cite{liu2024quantum} suggests a variational quantum approach to implement the Newton-Raphson iteration for the AC PF. Because the method relies on the LCU decomposition of the PF Jacobian, it is challenged by the curse of dimensionality. Reference~\cite{Kondo2022} proposes a novel quantum measurement protocol, wherein specific measurement matrices can be diagonalized using a few quantum-implementable unitaries. In~\cite{qopf}, we adopted this protocol to quantum observables associated with the AC optimal power flow (AC OPF) problem. Despite its resemblance to the AC OPF, the AC PF problem requires special treatment in terms of measuring quantum observables. 

To bridge the identified gaps, this work proposes a data-based quantum power flow framework to solve the AC PF problem. Its technical contributions are on three fronts: \emph{i)} Develop a model for solving the AC PF via a hybrid classical-quantum algorithm (Section~\ref{sec:qpf}); \emph{ii)} Leverage a data-embedded VQC and train a QML model to predict AC PF solutions in an unsupervised manner (Section~\ref{sec:data}); \emph{iii)} Expedite the computation of VQC/QML by reformulating PF to involve fewer expectations over quantum observables (Section~\ref{sec:coloring}). Numerical tests on the IEEE 14-bus system demonstrate that the novel QPF framework predicted AC PF solutions with smaller errors while using significantly fewer parameters than a deep neural network (Section~\ref{sec:tests}). Section~\ref{sec:conclude} concludes our findings and suggests opening research directions.


\section{Quantum Computing Preliminaries}\label{sec:quantum}
A quantum computer operating on $\log N$ quantum bits (qubits) is a random sampler of $\log N$ classical bits. Every time this computer is run, we read a string of $\log N$ classical bits at its output. Quantum computing relies on manipulating the probability mass function (PMF) of the $N$ bitstring outcomes, and is founded on four basic postulates~\cite{nielsen00}. According to the first, a quantum system operating on $\log N$ qubits is described by a \emph{state vector} $\bx\in\mathbb{C}^N$. The Dirac or \emph{ket} notation $\ket{\bx}$ highlights the fact that a quantum state vector is always of unit norm, so that $\|\bx\|_2^2=\bx^\mcH\bx=\braket{\bx|\bx}=\sum_{k=0}^{N-1}|x_k|^2=1$, where the \emph{bra} notation $\bra{\bx}$ denotes $\bra{\bx}=\bx^\mcH$. This vector captures the PMF related to the quantum system, in the sense that $|x_k|^2$ is the probability of sampling the bitstring corresponding to the binary representation of integer $k=0,\ldots,N-1$.

The second postulate predicates that the only operation to be applied on a quantum state is a unitary matrix $\bU\in\mathbb{C}^{N\times N}$ to transform state $\ket{\bx}$ to $\ket{\by}=\bU\ket{\bx}$. Not all unitaries can be efficiently implemented in quantum hardware. We can typically implement unitaries operating on one or two qubits. Some examples of single-qubit unitaries are:
\begin{equation}\label{eq:example}
\bX=\begin{bmatrix}0 & 1\\1& 0 \end{bmatrix} \quad \text{and}\quad 
\bR_X(\theta)=\begin{bmatrix}\cos(\theta) &-i\sin(\theta) \\-i\sin(\theta)& \cos(\theta) \end{bmatrix}.
\end{equation}
Matrix $\bX$ implements the quantum NOT gate, and $\bR_X(\theta)$ is an example of a parameterized unitary, where $\theta\in[0,2\pi]$ is controlled externally.

According to the third postulate, the output of a quantum state can be associated with a Hermitian matrix $\bH$ to provide the expectation over a \emph{quantum observable} as the quadratic function 
\[F=\braket{\bx|\bH|\bx}=\bx^\mcH \bH \bx.\]
If $\bH$ is diagonal, then obviously $F=\sum_{k=0}^{N-1}|x_k|^2H_{kk}$ where $H_{kk}$ are the diagonal entries of $\bH$. If a random variable takes values $H_{kk}$ with probability $|x_k|^2$ for all $k$, then $F$ is essentially the mean of that random variable.

Per the fourth postulate, if a quantum system is described by $\ket{\bx_1}$ and a second one by $\ket{\bx_2}$, the joint system is characterized by state $\ket{\bx_1,\bx_2}=\bx_1\otimes \bx_2$, where $\otimes$ is the Kronecker product.  

\begin{figure}[t]
\centering
\includegraphics[width=1\linewidth]{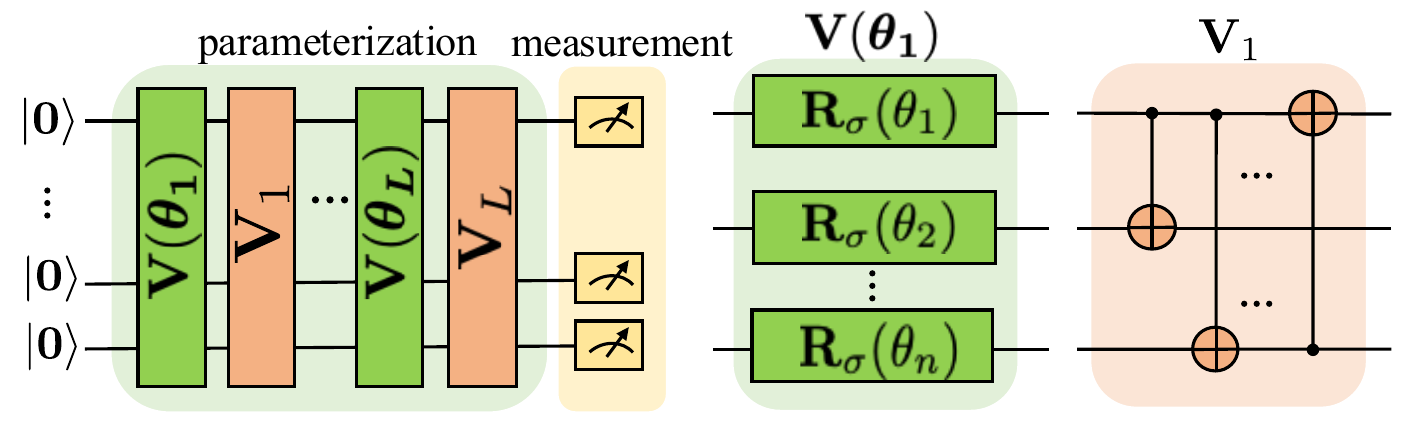}
\caption{A typical architecture of a VQC operating on an $\log N$ qubits. \emph{Left:} An $L$-layered hardware-efficient VQC~\cite[Ch.~5]{schuld2019}. \emph{Middle:} A parameterized layer of the hardware-efficient VQC, where $\bR_{\sigma}$ denotes single-qubit Pauli rotation gates $(\bR_X, \bR_Y, \bR_Z)$. \emph{Right:} One cyclic entanglement layer comprises $\log N$ two-qubit CNOT gates.}
\label{fig:vqc}
\end{figure}

The \emph{variational quantum eigensolver} (VQE) is a computational task tailored to NISQ systems. Given an $N\times N$ Hermitian $\bH_0$, VQE aims to find its minimum eigenvalue and associated eigenvector. Because $N$ can be large, VQE captures the sought eigenvector as a parameterized quantum state $\ket{\bpsi(\btheta)}$ using $\log N$ qubits, and solves the related \emph{variational} eigenproblem:
\begin{equation}\label{eq:vqe}
\min_{\btheta}~F_0(\btheta):=\braket{\bpsi(\btheta)|\bH_0|\bpsi(\btheta)}
\end{equation}
over a vector of parameters $\btheta\in\mathbb{R}^P$, where typically $P$ scales polynomially with $\log N$. The state $\ket{\bpsi(\btheta)}$ is prepared by feeding $\ket{\bzero}$ into a \emph{variational quantum circuit} (VQC) as
\[\ket{\bpsi(\btheta)}=\bV(\btheta)\ket{\bzero}\]
where $\ket{\bzero}$ is the first column of the identity $\bI_N$, and $\bV(\btheta)$ is a parameterized unitary describing the VQC. The VQC consists of layers of parameterized single-qubit gates and fixed two-qubit gates, shown in green and orange, respectively, in Fig.~\ref{fig:vqc}.

The VQE optimization in \eqref{eq:vqe} is solved in a hybrid classical/quantum fashion. The VQC evaluates the objective function $F_0(\btheta)$ and its gradient, vector $\btheta$ is updated using gradient descent (GD) updates on a classical computer. Critically, for a broad family of VQC architectures, the gradient $\nabla_{\btheta}F_0(\btheta)$ can be computed by evaluating $F_0$ at $2P$ additional points. In detail, according to the \emph{parameter shift rule} (PSR), a partial derivative can be evaluated exactly as the finite difference 
\begin{equation}\label{eq:psr}
 \frac{\partial F_0(\btheta)}{\partial \theta_p}=\tfrac{1}{2}\left(F_0(\btheta+\tfrac{\pi}{2}\be_p)-F_0(\btheta-\tfrac{\pi}{2}\be_p)\right)
\end{equation}
where $\be_p$ is the $p$-th column of the identity $\bI_P$. Building on VQE, we next explain how a single instance of the AC PF task can be solved using a variational quantum approach.

\section{Single-Instance Quantum Power Flow}\label{sec:qpf}
After briefly reviewing the AC-PF task, we reformulate it as an interesting generalization of the VQE problem. Consider a power system consisting of $N$ buses indexed by $n\in\{1,\ldots,N\}$. Let $p_n+jq_n$ and $v_n$ denote the complex power injection and the complex voltage phasor at bus $n$, respectively. Vector $\bv\in\mathbb{C}^N$ collects all voltage phasors. In its standard form, the AC power flow problem entails solving the AC PF equations to find $\bv$ given two specifications per bus. Depending on its specifications, a bus can belong to one of the following three types: \emph{i)} One generator bus is selected as the reference or slack bus, for which we specify $|v_n|$ and set its phase angle to zero; \emph{ii)} For the remaining generator buses (also known as PV buses), we specify $(p_n,|v_n|)$ or, equivalently, $(p_n,|v_n|^2)$. We will henceforth assume the latter for convenience. \emph{iii)} For load buses (also known as PQ buses), we specify $(p_n,q_n)$. Because the AC PF equations are invariant with respect to a phase shift in $\bv$, nulling the voltage phase angle of the reference bus is unnecessary. We are thus left with $S=2N-1$ real-valued specifications. Let $b_s$ denote the value of specification $s$, and collect specifications in $\bb\in\mathbb{R}^{S}$.

Solving the AC-PF problem is equivalent to solving the following set of $S$ quadratic equations over $\bv$~\cite{ZhuGia12, Madani2019}
\begin{equation}\label{eq:pf0}
\bv^\mcH \bH_s \bv=b_s, \quad s=1,\ldots,S
\end{equation}
where $\bH_s$ are known sparse Hermitian matrices that depend on the bus admittance matrix $\bY$. Grid operators solve \eqref{eq:pf0} routinely under different loading conditions and generator settings. As long as $\bY$ remains unaltered, an instance of the AC-PF is characterized by its specification vector $\bb$. 


For a particular PF instance $\bb$, a solution is typically found via iterative methods, such as the Gauss-Seidel, the impedance matrix, or the Newton-Raphson method. Alternatively, recent approaches train ML models to predict AC-PF solutions once presented with a $\bs$ at their input. The idea is to offload computational cost from real-time to offline, given that the ML model should be trained over a sufficiently representative dataset of AC-PF instances $\{\bb_t\}_{t=1}^T$. In lieu of a classical ML model, we suggest predicting AC-PF solutions using a VQC. Before dealing with multiple PF instances, let us first explain how to deal with a single PF instance. 

To this end, first note that rather than solving the quadratic equations in \eqref{eq:pf0}, a PF solution can be found as a minimizer of the unconstrained optimization
\begin{equation}\label{eq:pf1}
\min_{\bv \in \mathbb{C}^N}\sum_{s=1}^S \left(\bv^\mcH \bH_{s} \bv-b_s\right)^2.
\end{equation} 
This formulation is also relevant for solving the power system state estimation (PSSE) task, in which case the specifications $b_s$ correspond to noisy measurements of grid quantities in addition to $(p_n,q_n,|v_n|)$, such as apparent currents and line power flows, so that $S\geq 2N-1$. The objective in \eqref{eq:pf1} is quartic in $\bv$ and can be easily shown as non-convex; convex reformulations of \eqref{eq:pf1} for handling the PF and the PSSE tasks have been studied in~\cite{Madani2019,ZhuGia12,redux}.

Spurred by the VQE idea, it is tempting to capture directly $\bv$ by the state vector $\ket{\bpsi(\btheta)}$ of a VQC on $\log N$ qubits, presuming for now that the number of buses $N$ is a power of 2. This mapping does not work as $\bv$ is not of unit norm. In fact, because voltage magnitudes are typically in the range of one per unit (pu), we expect $\|\bv\|_2$ to be in the order of $\sqrt{N}$. As in~\cite{scaling_state}, we can introduce a classical optimization variable $\alpha>0$ and model the power system state variationally as
\begin{equation}\label{eq:scale}
\bv(\btheta)=\sqrt{\alpha}\ket{\bpsi(\btheta)}
\end{equation}
so that $\alpha=\|\bv(\btheta)\|_2^2$. If we define the expectation $F_s(\btheta):=\braket{\bpsi(\btheta)|\bH_s|\bpsi(\btheta)}$, a PF specification can be expressed in its variational form as
\begin{equation}\label{eq:fsvar}
(\bv(\btheta))^\mcH\bH_s\bv(\btheta)=\alpha\braket{\bpsi(\btheta)|\bH_s|\bpsi(\btheta)}=\alpha F_s(\btheta).
\end{equation}
Based on~\eqref{eq:fsvar}, we can attempt solving the AC-PF variationally over $(\btheta,\alpha)$ instead of $\bv$ as
\begin{equation}\label{eq:qpf}\tag{\texttt{QPF}} 
\min_{\btheta,\alpha>0}~f(\btheta,\alpha)=\sum_{s=1}^{S} \left(\alpha F_s(\btheta)- b_s \right)^2.
\end{equation}
However, unlike VQE, the objective in \texttt{(QPF)} is not a linear function of a single expectation over a quantum observable. It is rather the sum of a large number $S$ of \emph{squared} expectation values. These two issues render measuring the objective and its gradient on a VQC non-straightforward. To address these challenges, we leverage the special structure of \texttt{(QPF)} and introduce a novel quantum measuring protocol. 

The presentation of this measurement protocol is postponed to Section~\ref{sec:coloring}. For now, it suffices to assume that we can readily evaluate the gradients of $f(\btheta,\alpha)$ on a VQC, so that we can find a stationary point of \texttt{(QPF)} using GD updates
\begin{align*}
\btheta^{k+1}&=\btheta^{k}-\mu\nabla_{\btheta}f(\btheta^k,\alpha^k)\\
\alpha^{k+1}&=\max\left\{\alpha^{k}-\mu\frac{\partial f(\btheta^k,\alpha^k)}{\partial \alpha},0\right\}
\end{align*}
for a step size $\mu>0$.

We close this section with a note on zero-padding the grid state. If the size of the power system $N$ is a power of 2, the voltage vector $\bv$ can be modeled using $\log N$ qubits. Otherwise, we need $\lceil \log N \rceil$ qubits and zero-pad matrices $\bH_s$ so they become of size $2^{\lceil \log N \rceil}\times 2^{\lceil \log N \rceil}$. In the latter case, the $2^{\lceil \log N \rceil}$-dimensional quantum state $\ket{\bpsi(\btheta)}$ is partitioned into the first $N$ entries comprising subvector $\bpsi_1(\btheta)$, and the remaining ones comprising $\bpsi_2(\btheta)$. We now have that
\[\bv(\btheta)=\alpha\bpsi_1(\btheta) ~~~\text{and}~~~ 
F_s(\btheta)=\alpha (\bpsi_1(\btheta))^\mcH\bH_s \bpsi_1(\btheta).\]  However, scaling is not as straightforward as before because
\[\|\bv(\btheta)\|_2^2=\alpha\|\bpsi_1(\btheta)\|_2^2=\alpha(1-\|\bpsi_2(\btheta)\|_2^2)\]
given that $\ket{\bpsi(\btheta)}$ has unit norm. Subvector $\bpsi_2(\btheta)$ does not contribute to any expectation as it is multiplied by the zero-padded part of the augmented $\bH_s$, yet its norm may still be considerable. If $\|\bpsi_2(\btheta)\|_2$ is large, then $\|\bpsi_1(\btheta)\|_2$ gets small and challenges the quantum measurement process. To avoid solutions where a small $\|\bpsi_1(\btheta)\|_2$ is compensated by a large value of $\alpha$, we add the constraint $\alpha\leq 1.1^2N$. This is a safe upper bound on $\alpha$ as voltages are typically less than 1.1~pu.

\section{Data-Based Quantum Power Flow}\label{sec:data}
So far, we have explained how the VQC parameters $\btheta$ can be optimized to solve a single PF instance. In Section~\ref{sec:qpf}, the vector $\bb$ of PF specifications appeared only in the objective of \eqref{eq:qpf}. This section embeds PF specifications into the VQC state and trains the VQC parameters $\btheta$ based on multiple PF instances $\{\bb_t\}_{t=1}^T$. In classical ML models trained to predict PF solutions, vector $\bb$ is treated as the input or feature vector. Analogously, our goal is to make the VQC state dependent on data and parameters alike as $\ket{\bpsi(\bb,\btheta)}$.

\begin{figure}[t]
\centering
\includegraphics[width=0.85\linewidth]{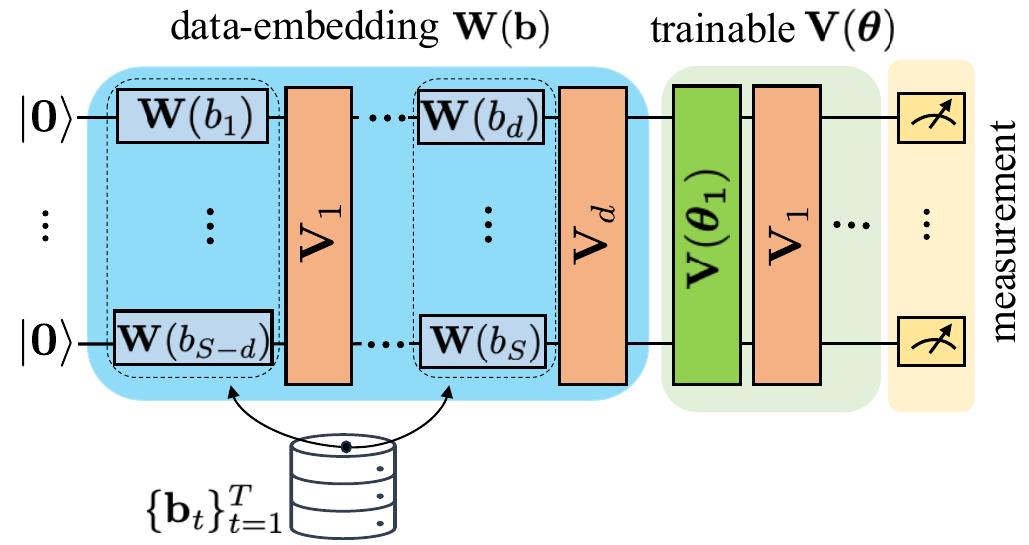}
\vspace*{-1em}
\caption{A data-embedding VQC includes three main blocks: the data embedding block, the trainable block, and the measurement block. Data $\bb \in \mathbb{R}^S$ can be embedded as parameters of $S$ parameterized single-qubit gates (colored in blue), arranged into $d=S/\log N$ layers.}
\label{fig:dataVQC}
\end{figure}

To accomplish this goal, we propose using a \emph{data-embedded VQC}; see~\cite[Ch.~5]{schuld2021}. This type of VQC comprises three blocks, as shown in Fig.~\ref{fig:dataVQC}: a data-embedding block, a trainable block, and the measurement layers. The last two are identical to those described in Section~\ref{sec:qpf}. In the QML context, the trainable parameters $\bb$ are analogous to the weights of classical ML models. The data embedding block inputs data into the QML model. Similarly to VQC weights $\btheta$, VQC data $\bb$ are loaded into the VQC as parameters of another unitary $\bW(\bb)$. 

With two blocks of parameterized unitaries, the VQC state is now described as 
\[\ket{\bpsi(\btheta, \bb)}=\bV(\btheta)\bW(\bb)\ket{\bzero}\]
so the expectation $F_s(\btheta,\bb)=\braket{\bpsi(\btheta, \bb)|\bH_s|\bpsi(\btheta, \bb)}$ becomes a nonlinear function of $\btheta$ and $\bb$ as with DNNs~\cite{schuld2021effect}. Since this function is trigonometric, PF data are normalized within $[0,2\pi]$. Both $\bV(\btheta)$ and $\bW(\bb)$ operate on $\log N$ qubits. To achieve this, we partition $\bb$ into $\log N$ subvectors, each of dimension $d=\lceil S/\log N \rceil$. Each subvector is embedded into $d$ gates operating on the same qubit, as shown in Fig.~\ref{fig:dataVQC}. The entire data-embedding block can be repeated several times to enhance the representation capability of the VQC~\cite{schuld2021effect}. If $S$ is not divisible by $\log N$, vector $\bb$ is padded with zeros. Note that parameterized single-qubit gates, such as $\bR_X(\theta)$ in \eqref{eq:example}, when fed with $0$, become identity operators and thus act trivially on qubits. 

Having embedded PF specifications into the VQC state, the VQC can now be trained in an unsupervised fashion to learn the solutions of multiple PF instances. More specifically, the optimal VQC parameters $(\btheta,\alpha)$ can be found by minimizing the ensuing loss function using gradient descent:
\begin{equation}\label{eq:loss}
\min_{\btheta,\alpha>0}~\sum_{t=1}^T \sum_{s=1}^{S} \left(\alpha F_s(\btheta,\bb_t)- b_t^s \right)^2
\end{equation}
where $b_t^s$ is the $s$-th entry of $\bb_t$. Mini-batch gradients can be implemented by sampling a few instances $\bb_t$ per iteration~\cite{sweke2020}.

\section{Efficient Quantum Power Flow Measurements}\label{sec:coloring}
In this section, we propose a novel approach for efficiently measuring the PF objective and its gradient using a VQC. The proposed protocol applies to QPF on a per-instance basis, so we drop dependence on $\bb_t$. Upon ignoring inconsequential constants, the cost in \eqref{eq:qpf} can be expanded as
\begin{equation}\label{eq:fun}
f(\btheta,\alpha)=\alpha^2\sum_{s=1}^{S}F_s^2(\btheta)- 2\alpha\sum_{s=1}^S b_s F_s(\btheta).
\end{equation}
Leaving the $\alpha$ factor aside, the second summand can be expressed as a single expectation
\begin{equation}\label{eq:G}
G(\btheta)=\braket{\bpsi(\btheta)|\bG|\bpsi(\btheta)}~~\text{where}~~\bG=\sum_{s=1}^S b_s\bH_s.    
\end{equation}

Can we efficiently measure $G(\btheta)$? One option would be to express $\bG$ as a linear combination of unitaries (LCU)~\cite[p.~138]{schuld2021}. An LCU-based approach is practical only if the unitaries can be efficiently implemented on a quantum computer and their number scales polynomially in $\log N$. We conjecture that neither of these conditions is met. 

Another idea would be trying the eigenvalue decomposition of $\bG=\bU_g\bLambda_g\bU_g^\mcH$. We can apply the unitary $\bU_g^\mcH$ on the VQC state and measure the derived state $\bU_g^\mcH\ket{\bpsi(\btheta)}$ using the diagonal matrix $\bLambda_g$. This approach requires computing an eigenvalue decomposition per PF instance, and matrix $\bU_g^\mcH$ is unlikely to be efficiently realizable.

In our previous work~\cite{qopf}, we adapted the so-called extended Bell measurement (XBM) protocol of \cite{Kondo2022} to efficiently measure observables to solve the OPF on a VQC. We briefly review the key ideas and extend them to the PF setting. 

\begin{lemma}[\cite{Kondo2022}]\label{le:coloring}
Given the admittance matrix $\bY$ of a power system, there exists a set of unitary matrices $\{\bU_i\}_{i=1}^C$ so that the specification matrices can be decomposed as
\[\bH_s=\sum_{i=1}^C \bU_i\bLambda_i^s\bU_i^\mcH~~\text{for}~~s=1,\ldots,S\]
where 
\begin{enumerate}
    \item[\emph{i)}] Matrices $\bLambda_i^s$ are diagonal, real-valued, and can be easily computed. The number of non-zero entries scales with the maximum degree of the power system graph, that is $\mcO(1)$; and
    \item[\emph{ii)}] The unitary matrices $\bU_i^\mcH$ are qubit efficient, i.e., they can be realized using at most $\log N + 1$ single- and two-qubit gates.
\end{enumerate}
\end{lemma}

The critical feature of Lemma~\ref{le:coloring} is that all $S=2N-1$ specification matrices can be decomposed using the same $C$ unitaries. Numerical tests using all power system benchmarks in the \texttt{pglib} database indicate that the number of unitaries $C$ scales polynomially in $\log N$. 

Based on Lemma~\ref{le:coloring}, matrix $\bG$ can be decomposed as
\begin{align*}
\bG&=\sum_{s=1}^S b_s\bH_s=\sum_{s=1}^S b_s\left(\sum_{i=1}^C\bU_i\bLambda_i^s\bU_i^\mcH\right)\\
&=\sum_{i=1}^C\bU_i\underbrace{\left(\sum_{s=1}^S b_s \bLambda_i^s\right)}_{:=\bLambda_i(\bb)}\bU_i^\mcH=\sum_{i=1}^C\bU_i\bLambda_i(\bb)\bU_i^\mcH.
\end{align*}
Thanks to this decomposition, the expectation in \eqref{eq:G} can be measured as the sum of $C$ expectation values:
\begin{align*}
G(\btheta)&=\sum_{i=1}^C \braket{\bpsi(\btheta)| \bU_i\bLambda_i(\bb)\bU_i^{\mcH}|\bpsi(\btheta)}\\
&=\sum_{i=1}^C \underbrace{\braket{\bpsi_i(\btheta)| \bLambda_i(\bb)|\bpsi_i(\btheta)}}_{:=G_i(\btheta;\bb)}=\sum_{i=1}^C G_i(\btheta;\bb).
\end{align*}
To measure $G_i(\btheta;\bb)$, apply unitary $\bU_i^\mcH$ on the VQC state to generate $\ket{\bpsi_i(\btheta)}=\bU_i^\mcH\ket{\bpsi(\btheta)}$, and measure the latter state on $\bLambda_i(\btheta)$. The expectations $G_i(\btheta;\bb)$ can be measured using a serial or parallel architecture. The parallel one consists of $C$ replicas of the original VQC, and applies a different unitary at the end of each replica. The serial architecture operates on a single VQC replica and appends a different unitary each time, iterating over $C$. The parallel architecture is $C$ times faster but requires $C$ times more quantum hardware.

Let us focus on the first summand in \eqref{eq:fun}. Although this term does not depend on PF specifications, it is a quartic function of $\ket{\bpsi(\btheta)}$. To address this, we propose using two replicas of the VQC to generate the joint quantum state
\[\ket{\tbpsi(\btheta)}=\ket{\bpsi(\btheta)}\otimes \ket{\bpsi(\btheta)}.\]
Note that although the two VQCs have identical states, when sampled, they produce different binary strings due to the random nature of the quantum circuit. The two states are identical because the two VQCs have the same topologies and are fed with the same parameters $\btheta$. 

Dropping the $\alpha^2$ factor, the dependence on $\btheta$, and the ket notation for simplicity, we write the first summand in \eqref{eq:fun} as
\begin{align*}
\tilde{G}&=\sum_{s=1}^S F_s^2=\sum_{s=1}^S (\bpsi^\mcH\bH_s\bpsi)(\bpsi^\mcH\bH_s\bpsi)\\
&=\sum_{s=1}^S (\bpsi\otimes \bpsi)^\mcH(\bH_s\otimes \bH_s)(\bpsi\otimes \bpsi)\\
&=\tbpsi^\mcH \underbrace{\left(\sum_{s=1}^S \bH_s\otimes \bH_s \right)}_{:=\tbG}\tbpsi=\tbpsi^\mcH \tbG\tbpsi.
\end{align*}
Therefore, the term $\tilde{G}(\btheta)$ is an expectation over the joint state $\ket{\tbpsi(\btheta)}$ with associated Hermitian matrix $\tbG$.

Leveraging Lemma~\ref{le:coloring}, matrix $\tbG$ can be decomposed as 
\begin{align*}
\tbG&=\sum_{s=1}^S \left(\sum_{i=1}^C \bU_i\bLambda_i^s\bU_i^\mcH\right) \otimes \left(\sum_{j=1}^C \bU_j\bLambda_j^s\bU_j^\mcH\right)\\
&=\sum_{i=1}^C\sum_{j=1}^C \left(\bU_i\otimes \bU_j\right)^\mcH \underbrace{\left(\sum_{s=1}^S\bLambda_i^s\otimes \bLambda_j^s\right)}_{=\bLambda_{ij}}\left(\bU_i\otimes \bU_j\right)\\
&=\sum_{i=1}^C\sum_{j=1}^C \left(\bU_i\otimes \bU_j\right)^\mcH \bLambda_{ij}\left(\bU_i\otimes \bU_j\right).
\end{align*}

\begin{algorithm}[t]
	\caption{Measure expectation $\tilde{G}(\btheta)$}
    \label{alg:tG}
    \begin{flushleft}
	\begin{algorithmic}[1]
		\renewcommand{\algorithmicensure}{\textbf{Initialize:}} 
		\ENSURE $\tilde{G}(\btheta)=0$.
		\FOR{$i=1$ to $C$}
        \FOR{$j=1$ to $C$}
        \STATE Apply $\bU_i^\mcH$ on first VQC to generate $\ket{\bpsi_i(\btheta)}=\bU_i^\mcH\ket{\bpsi(\btheta)}$.
		\STATE Apply $\bU_j^\mcH$ on second VQC to generate $\ket{\bpsi_j(\btheta)}=\bU_j^\mcH\ket{\bpsi(\btheta)}$.
		\STATE Measure $\ket{\bpsi_i(\btheta)}\otimes \ket{\bpsi_j(\btheta)}$ on diagonal $\bLambda_{ij}$.
        \STATE Add measurement to $\tilde{G}(\btheta)$.
        \ENDFOR
        \ENDFOR
	\end{algorithmic}
    \end{flushleft}
\end{algorithm}
Based on this decomposition, the term $\tilde{G}(\btheta)$ can be measured using Algorithm~\ref{alg:tG}. Because we need to measure all unique pairs of unitaries, the two VQCs are recompiled $C(C+1)/2$ times. It is important to note that measuring pairs of unitaries is required only during training.

\section{Numerical Tests}\label{sec:tests}

\begin{figure}[t]
\centering
\includegraphics[width=0.75\linewidth]{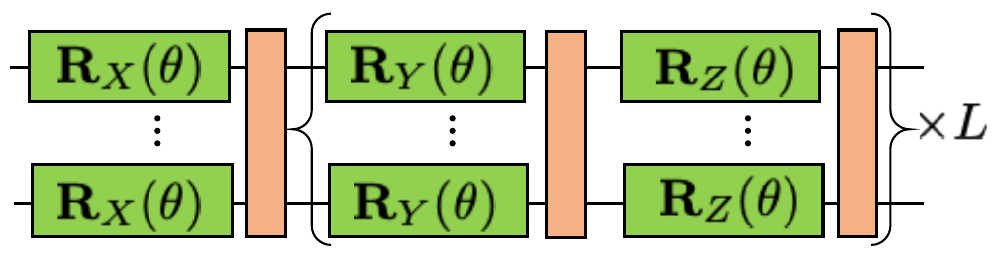}
\caption{VQC architecture used in numerical tests. Blocks in orange represent cyclic entanglement with CNOT gates. For the single-instance and multi-instance QPFs, we used $L=3$ and $L=6$ layers, respectively.}
\label{fig:vqc_test}
\end{figure}

The proposed QPF framework was evaluated using the IEEE 14-bus system with $S=27$ specifications. To generate multiple PF instances, benchmark PF specifications were perturbed by zero-mean, white Gaussian noise of standard deviation of 0.05~pu for voltage magnitude specifications, and 20\% of the nominal values for power injections. VQCs were coded in Python using Pennylane's exact quantum simulator~\cite{Pennylane}. The VQC operated on 4 qubits with matrices zero-padded to reach the dimension of 16. All tests used the VQC architecture of Fig.~\ref{fig:vqc_test}. VQC weights $\btheta$ were initialized so the related quantum state corresponds to the flat voltage profile, using the method of~\cite{BravoPrieto2023variationalquantum}. Variable $\alpha$ was initialized to $\sqrt N$. Gradient descent iterations were terminated when the Euclidean norm of the gradient was smaller than 0.01.  

\begin{figure}[t]
\centering
\includegraphics[width=0.87\linewidth]{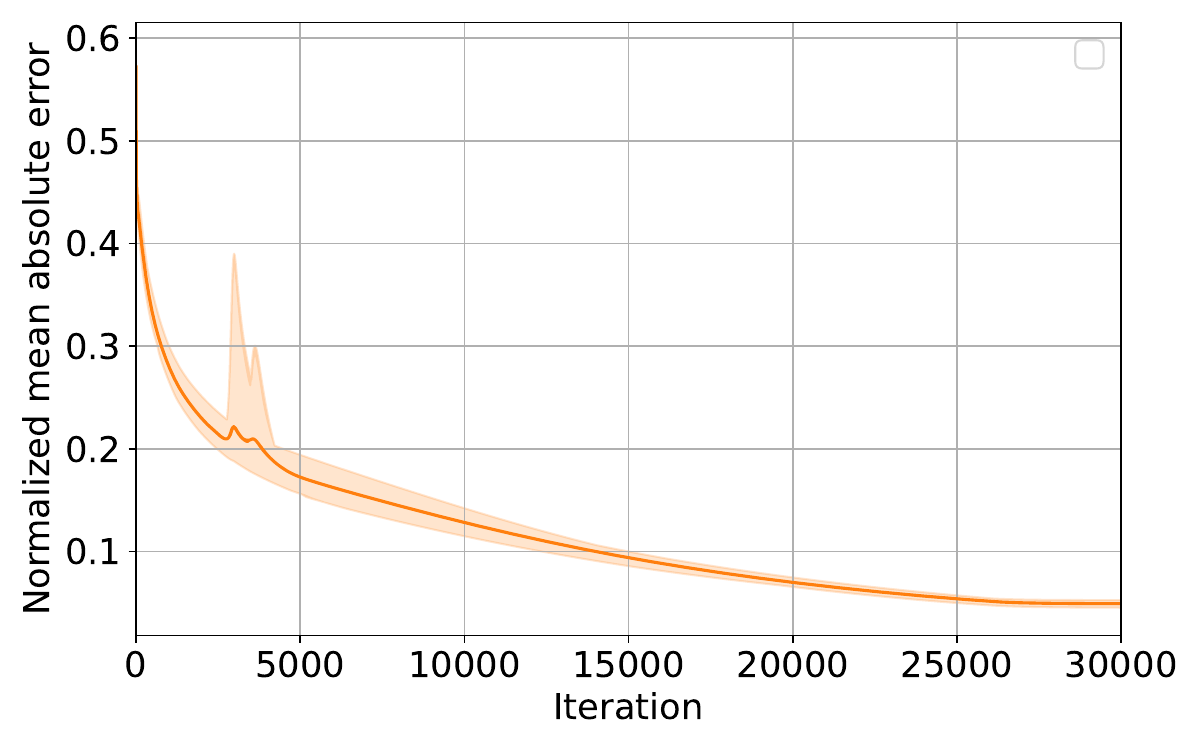}
\vspace*{-1em}
\caption{Convergence of the normalized mean absolute error (NMAE) for the single-instance QPF across gradient descent iterations. The plot shows the confidence intervals within one standard deviation around the mean, computed over $10$ PF problem instances.}
\label{fig:deterministic}
\end{figure}

We first tested the performance of the single-instance OPF in terms of the normalized mean absolute error (NMAE) $\|\hbb-\bb\|_1/\|\bb\|_1$, where vector $\hbb$ evaluates the PF specifications based on the obtained voltage vector. We ran the single-instance QPF on 10 instances and $P=28$ trainable parameters. Step sizes were set as $\mu_{\theta}$ and $\mu_{\alpha}$ were set to $5\times10^{-5}$ across iterations. Figure~\ref{fig:deterministic} shows that the relative error $\|\hbb-\bb\|_1/\|\bb\|_1$ converges to a point around $5\%$ with small deviation. This demonstrates that QPF found near-optimal PF solutions across all 10 instances.

\begin{figure}[t]
\centering
\includegraphics[width=0.87\linewidth]{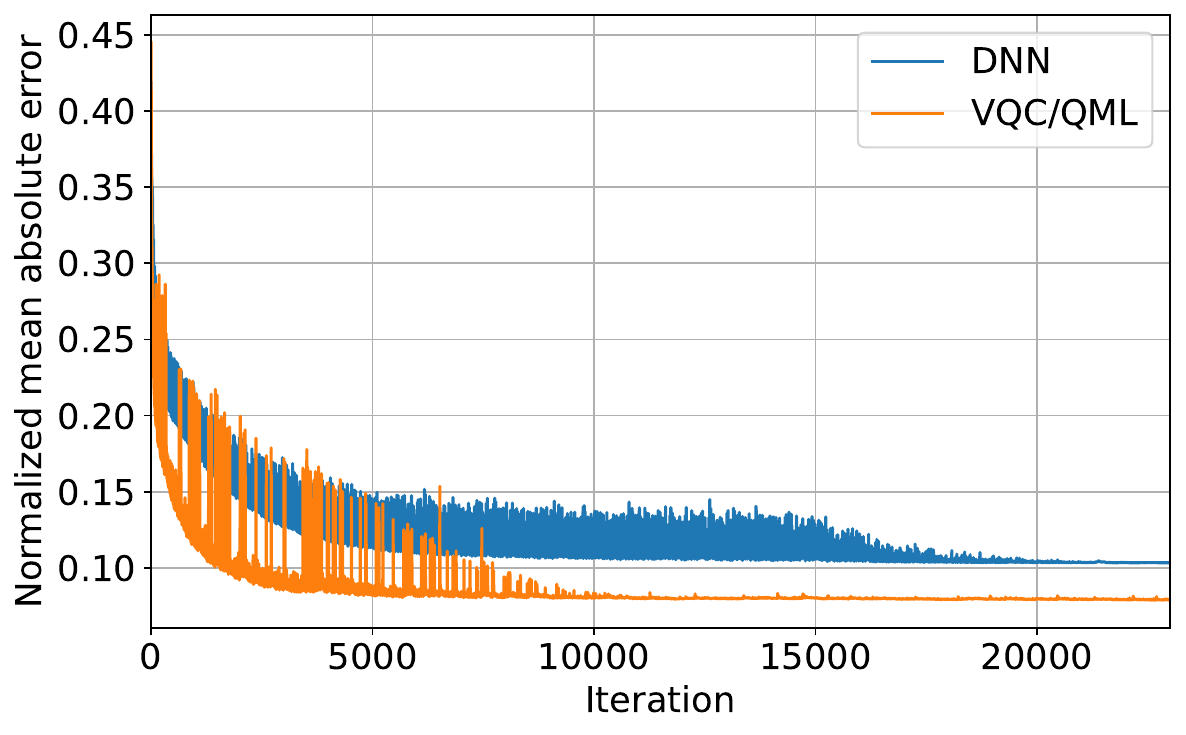}
\vspace*{-1em}
\caption{Convergence of NMAE attained by VQC/QML and DNN across gradient descent iterations.}
\label{fig:sampling}
\end{figure}

\begin{figure}[t]
\centering
\includegraphics[width=0.87\linewidth]{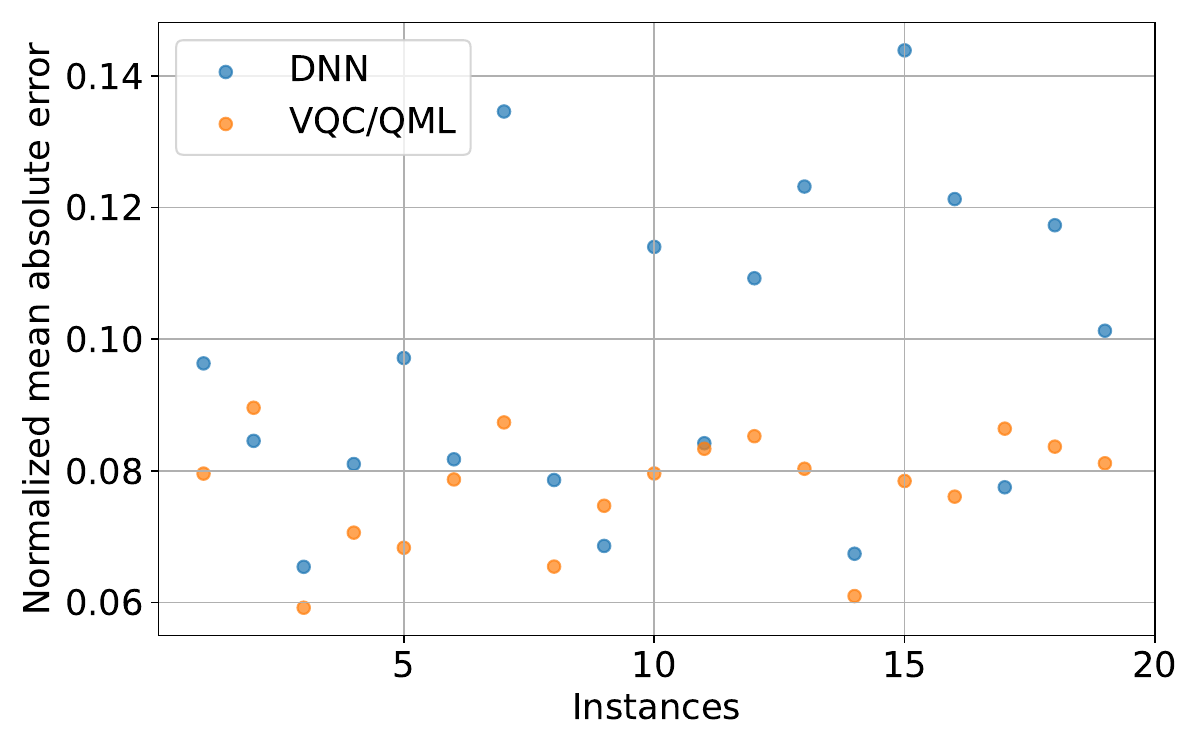}
\vspace*{-1em}
\caption{Comparing the VQC/QML and DNN regarding their NMAE over 20 testing AC-PF instances.}
\label{fig:inference}
\end{figure}

We subsequently trained a QML model to learn multiple PF solutions and compared it against a classical DNN model. The DNN architecture was selected based on four-fold cross-validation on 80 PF problem instances upon testing different step sizes. The number of input layers was set to the length of $\bb_t$. We used two fully connected hidden layers of dimension 10 each. The DNN outputs the real and imaginary parts of voltages of a total dimension of 32. Rectified linear units (ReLU) were used for all layers. Overall, the chosen DNN architecture used 690 weights. Regarding QML, we used the VQC architecture of Fig.~\ref{fig:vqc_test} with $L=6$ layers, resulting in only $P=52$ trainable weights. Gradient descent step sizes for DNN and VQC/QML were adjusted per the exponentially decaying rule as $0.9999^{k}$ and $0.9995^{k}$ at iteration $k$, respectively. Figure~\ref{fig:sampling} shows that the VQC/QML achieved faster convergence and smaller NMAE than DNN. Correspondingly, the VQC/QML model attained smaller prediction errors than DNN in 17 out of 20 instances, as shown in Fig.~\ref{fig:inference}. This test corroborates that VQC/QML attained better prediction errors and faster training than DNN, while using an order of magnitude fewer trainable parameters.

\color{black}

\section{Conclusions}\label{sec:conclude}
To cope with the challenges of solving multiple PFs promptly, this work has explored training a VQC/QML model to predict AC PF solutions. PF specifications are embedded as VQC parameters, allowing data-based VQC/QML to be trained unsupervised. To expedite gradient computations of VQC/QML, the PF problem has been judiciously reformulated via a few expectations, whose gradients can be measured efficiently on VQCs. Numerical tests on the IEEE 14-bus system have shown that the proposed data-based QPF achieved improved prediction error while using fewer trainable parameters (weights) than a classical DNN. Capitalizing on these results, several exciting directions are opened for future research: \emph{d1)} skip the unitary pairs in training VQC/QML; \emph{d2)} design grid-informed VQC topologies; and \emph{d3)} explore other VQC data encoding strategies.

\balance
\bibliographystyle{IEEEtran}
\bibliography{myabbrv,sgc25,power,kekatos}
\end{document}